# Covariate adjustment in subgroup analyses of randomized clinical trials: A propensity score approach

**Authors:** Siyun Yang (MS), Fan Li[1] (PhD), Laine E. Thomas (PhD), Fan Li[2] (PhD)

**Correspondence to** Fan Li[2], Department of Biostatistics, Yale University School of Public Health, New Haven, Connecticut (email: fan.f.li@yale.edu )

**Author affiliations:**

Department of Biostatistics and Bioinformatics, Duke University School of Medicine, Durham, North Carolina (Siyun Yang, MS; Laine E. Thomas, PhD); Department of Statistical Science, Duke University, Durham, North Carolina (Fan Li[1], PhD); Department of Biostatistics, Yale University School of Public Health, New Haven, Connecticut (Fan Li[2], PhD); Duke Clinical Research Institute, Durham, North Carolina (Laine E. Thomas, PhD)

**Acknowledgement:**

The authors are grateful to the Society of Clinical Trials, Student Scholarship Committee for being selected as a finalist for the 2021 Society for Clinical Trials Thomas Chalmers Student Scholarship. We also thank Dr. Robert J. Mentz from Duke Clinical Research Institute for valuable comments which improved this manuscript.

**Conflict of interest:** none declared.

**Funding:** This research is supported in part by the Patient-Centered Outcomes Research Institute (PCORI) contract ME-2018C2-13289. The contents of this article are solely the responsibility of the authors and do not necessarily represent the view of PCORI.

**Data availability statement:** The HF-ACTION data that is used as an illustrative example in this study can be accessed through NIH at https://biolincc.nhlbi.nih.gov/studies/hf_action/

**Running head:** Propensity score weighting in subgroup analyses of RCTs

**Word Count:** (Abstract 274); (Main text 4111)

**Abbreviations:**

ANCOVA: analysis of covariance

RCT: randomized controlled trials

**Abstract**

**Background**: Subgroup analyses are frequently conducted in randomized clinical trials (RCT) to assess evidence of heterogeneous treatment effect across patient subpopulations. Although randomization balances covariates within subgroups in expectation, chance imbalance may be amplified in small subgroups and adversely impact the precision of subgroup analyses. Covariate adjustment in overall analysis of RCT is often conducted, via either analysis of covariance (ANCOVA) or propensity score weighting, but covariate adjustment for subgroup analysis has been rarely discussed. In this article, we develop propensity score weighting methodology for covariate adjustment to improve the precision and power of subgroup analyses in RCTs.

**Methods**: We extend the propensity score weighting methodology to subgroup analyses by fitting a logistic regression propensity model with pre-specified covariate-subgroup interactions. We show that, by construction, overlap weighting exactly balances the covariates with interaction terms in each subgroup. Extensive simulations were performed to compare the operating characteristics of unadjusted estimator, different propensity score weighting estimators and the ANCOVA estimator. We apply these methods to the HF-ACTION trial to evaluate the effect of exercise training on 6-minute walk test in several pre-specified subgroups.

**Results**: Standard errors of the adjusted estimators are smaller than those of the unadjusted estimator. The propensity score weighting estimator is as efficient as ANCOVA, and is often more efficient when subgroup sample size is small (e.g.<125), and/or when outcome model is misspecified. The weighting estimators with full-interaction propensity model consistently outperform the standard main-effect propensity model.

**Conclusion**: Propensity score weighting is a transparent and objective method to adjust chance imbalance of important covariates in subgroup analyses of RCTs. It is crucial to include the full covariate-subgroup interactions in the propensity score model.

**Keywords**: Covariate adjustment; Power; Propensity score; Overlap weighting; Randomized controlled trials; Subgroup analyses.

**Background**

Subgroup analyses are frequently conducted in randomized clinical trials (RCTs) to assess evidence of heterogeneous treatment effect across patient subpopulations. When planned and interpreted properly, subgroup analyses offer valuable information about which population receives the most benefit or may be adversely impacted by an intervention, and can inform important decisions on personalized healthcare. While various data-driven methods have been developed to search for heterogeneous treatment effect,[1, 2] the vast majority of clinical trials still report confirmatory subgroup analyses of pre-specified subpopulations.[3] A major challenge for these analyses is a lack of power and precision,[4-6] which is amplified by chance imbalances that frequently occur in small subgroups. Optimizing precision is particularly important for subgroups because precision is already limited by smaller sample sizes and small interaction effects.[4, 5] In this article, we develop an easy-to-use propensity score weighting approach to improve the power of subgroup analyses by eliminating chance imbalances.

The standard confirmatory subgroup analysis of an RCT derives treatment effect estimates within pre-specified subgroups, and tests for interactions between those subgroups that partition the population. While randomization balances baseline covariates within each subgroup in expectation, chance imbalance is likely to occur within subgroups due to limited sample sizes and multiple opportunities for imbalance. Recent studies have demonstrated that covariate adjustment addresses baseline imbalance and improves the precision for analyzing RCTs with continuous and binary outcomes,[7-11] and time-to-event outcomes.[12] While methods for covariate adjustment are extensively studied for estimating the overall treatment effect, discussions on estimating the subgroup average treatment effect or heterogeneous treatment effect have been limited.

For estimating the average treatment effect with continuous outcomes, the two main approaches for covariate adjustment are analysis of covariance (ANCOVA) and propensity score weighting. Under mild conditions, Yang and Tsiatis,[13] and Tsiatis et al.[14] showed that the ANCOVA estimator with covariate-by-treatment interactions achieves the semiparametric efficiency for the average treatment effect. Shen et al.[15] demonstrated that the inverse propensity score weighting estimator under a logistic regression propensity score model is asymptotically equivalent to the efficient ANCOVA estimator. More recently, Zeng et al.[16] generalized this result to a large family of propensity score weighting estimators, including overlap weighting .[17-19] Despite the asymptotic equivalence, the two methods can differ in finite sample performance. Moreover, propensity score weighting has two advantages over ANCOVA: first, it avoids modeling the outcome data and is thus arguably more objective; second, it preserves the marginal treatment effect estimand for non-continuous outcomes. However, to our knowledge, neither ANCOVA nor propensity score weighting has been discussed in the context of covariate adjustment for subgroup analyses in RCTs. This paper fills the gap; we extend the ANCOVA model and consider two propensity score weighting methods[15, 20-22], inverse probability weighting and overlap weighting, for subgroup analyses. We perform extensive simulations to compare the operating characteristics of the propensity score weighting estimators with ANCOVA, and clarify their relative merits and limitations for subgroup analyses. We demonstrate the application of these approaches by estimating subgroup average treatment effects in the HF-ACTION study.[23]

**Causal Estimand and ANCOVA in Subgroup Analysis**

Consider a two-arm RCT of $N$ individuals. Let the binary treatment $Z_i = 1$ if unit $i$ receives the treatment and $Z_i = 0$ control. Under the potential outcome framework,[24] each individual has two potential outcomes, $Y_i(1)$ and $Y_i(0)$, corresponding to the treatment and control condition. Of the two, only the one corresponding to the actual treatment assigned is observed, denoted as $Y_i = Z_i Y_i(1) + (1 - Z_i)Y_i(0)$. We assume that $\boldsymbol{X_i} = (X_{i1}, \dots, X_{iJ})^T$ is a pre-specified list of $J$ covariates to be adjusted in the analysis. We also define a vector of $R$ pre-specified subgroup indicators, $\boldsymbol{S_i} = (S_{i1}, \dots, S_{iR})^T$. The subgroup variables *a priori* define the subgroup membership so that $S_{ir} = 1$ if unit $i$ belongs to the $r^{th}$ subgroup, and one unit can belong to multiple subgroups simultaneously. For generality, the subgroup indicator $\boldsymbol{S_i}$ may be a subset of $\boldsymbol{X_i}$, a function of $\boldsymbol{X_i}$, or not included in the set of $\boldsymbol{X_i}$ (e.g. in our HF-ACTION data example, subgroup variables such as Angina class, ventricular conduction do not belong to the set of pre-specified four adjustment variables). The $r^{th}$ subgroup average treatment effect is defined as:

$$\tau_r = E[Y(1) - Y(0)|S_r = 1]. \tag{1}$$

In RCTs, investigators often wish to test whether there is heterogeneous treatment effect across subgroup levels in one-at-a-time fashion for each $r$. In particular, the group comparisons are defined as $S_{ir} = 0$ versus $S_{ir} = 1$, while averaging over the levels of $\{S_{i1}, \dots, S_{iR}\} \setminus \{S_{ir}\}$. Therefore, the heterogeneous treatment effect across two subgroup levels can be formalized as

$$\delta_r^{HTE} = E[Y(1) - Y(0)|S_r = 1] - E[Y(1) - Y(0)|S_r = 0] \tag{2}$$

Assuming each unit $i$ is randomly assigned to the treatment or the control arm with a constant probability $0 < p < 1$. Because randomization balances observed and unobserved covariates in both the overall and subgroup samples in expectation, the following within-subgroup unadjusted difference-in-means estimator for subgroup average treatment effect is unbiased.

$$\hat{\tau}_r^{UNADJ} = \frac{\sum_{i=1}^N Y_i Z_i S_{ir}}{\sum_{i=1}^N Z_i S_{ir}} - \frac{\sum_{i=1}^N Y_i (1-Z_i) S_{ir}}{\sum_{i=1}^N (1-Z_i) S_{ir}} \tag{3}$$

However, due to the relatively small subgroup sample sizes, there is a higher chance for imbalance to occur within a subgroup, and adjusting for these imbalances may improve precision and power.[6-10, 25] For estimating the average treatment effect with a continuous outcome, the standard approach fits an ANCOVA model. We extend Yang and Tsiatis,[13] Tsiatis et al.[14] and Lin's[26] approach to estimate the subgroup average treatment effect, by separately fitting the ANCOVA model within each subgroup to allow for subgroup-specific coefficients. This is equivalent to fitting a full interaction model on the overall sample, whose predictors include adjustment variables $X_i$, treatment indicator, and their interactions with the subgroup variables $S_i$. We call this the "ANCOVA-S" model:

$$Y_i = \beta_0 + \boldsymbol{\beta}_1^T X_i + \boldsymbol{\beta}_2^T S_i + \beta_3 Z_i + \boldsymbol{\beta}_4^T (X_i S_i) + \boldsymbol{\beta}_5^T (Z_i S_i) + \boldsymbol{\beta}_6^T (X_i Z_i) \\ + \boldsymbol{\beta}_7^T (X_i S_i Z_i) + \epsilon_i, \quad i = 1, \ldots, N, \tag{4}$$

where $\beta = (\beta_0, \boldsymbol{\beta}_1^T, \boldsymbol{\beta}_2^T, \beta_3, \boldsymbol{\beta}_4^T, \boldsymbol{\beta}_5^T, \boldsymbol{\beta}_6^T, \boldsymbol{\beta}_7^T)$ are the regression coefficients, $(X_i S_i), Z_i S_i, (X_i Z_i), (X_i S_i Z_i)$ are the vectors of all 2-way and 3-way interactions between the subgroup indicators $S_i$, adjustment covariates $X_i$ and treatment assignment $Z_i$. The random error $\epsilon_i$ is assumed to have mean zero. Denote $\hat{Y}_i(1)$ and $\hat{Y}_i(0)$ as the predicted values from model (4) when setting $Z_i = 1$ and $Z_i = 0$, respectively. We define the "ANCOVA-S" estimator for subgroup average treatment effect as

$$\hat{\tau}_r^{ANCOVA} = \frac{\sum_{i=1}^{N}(\hat{Y}_i(1) - \hat{Y}_i(0))S_{ir}}{\sum_{i=1}^{N} S_{ir}} \qquad (5)$$

This estimator, $\hat{\tau}_r^{ANCOVA}$, is consistent for $\tau_r$ even if the functional forms of $X_i$ is incorrectly modeled, due to the similar arguments as in estimating the overall treatment effect.[13, 14] Compared to the $\hat{\tau}_r^{UNADJ}$ in (3), $\hat{\tau}_r^{ANCOVA}$ is more efficient when the adjusted covariates $X_i$ are predictive of $Y_i$. Variance and confidence intervals of $\hat{\tau}_r^{ANCOVA}$ can either be obtained using the delta method or bootstrap resampling.

**Propensity Score Weighting in Subgroup Analysis**

One limitation of ANCOVA is the potential for inviting a "fishing expedition", that is, one may search for a model specification that provides the largest overall and/or subgroup treatment effects and thus jeopardize the objectivity of statistical analysis.[14, 15] propensity score weighting—an approach that does not require access to the outcome data— is an attractive alternative approach to covariate adjustment;[16, 22] it achieves the same goal of improving precision, but offers better transparency by encouraging pre-specification and diagnostic assessment of covariate balance.

For subgroup analysis, we define the propensity score,[27] $e(\boldsymbol{X}, \boldsymbol{S}) = \Pr(Z = 1 | \boldsymbol{X}, \boldsymbol{S})$, as the conditional probability of receiving treatment given adjustment variables $\boldsymbol{X}$ and subgrouping variables $\boldsymbol{S}$. When the subgrouping variables $\boldsymbol{S}$ is an explicit function of $\boldsymbol{X}$, $\boldsymbol{S}$ can theoretically be omitted from the conditioning set. However, explicitly writing out $\boldsymbol{S}$ in this definition helps clarify the role of propensity scores for subgroup analysis.[28, 29] In RCTs, the treatment allocation

process is completely controlled and the true propensity score equals to the randomization probability of a specific trial (i.e. $e(X, S) = 0.5$ for all adjustment and subgrouping variables under a balanced treatment allocation design). This is distinct from observational studies, where the true propensity score is unknown and needs to be estimated from the data. Among the propensity score methods, Li et al.[17] showed incorporating different balancing weights leads to different target estimand and population in observational studies. In contrast, in RCTs the target estimand of all balancing weights reduces to the same estimand subgroup average treatment effect (see Supplementary Material A for technical details). Moreover, extreme weights rarely occur due to randomization. Even so, the use of an estimated propensity score will help to balance covariate distribution and improve finite sample properties of RCTs.

Similar to the causal subgroup analysis in observational studies,[29] we use the Hájek estimator, weighted difference-in-means in subgroup $r$, for estimating subgroup average treatment effect in RCTs:

$$\hat{\tau}_r^h = \frac{\sum_{i=1}^N Y_i Z_i w_{i1} S_{ir}}{\sum_{i=1}^N Z_i w_{i1} S_{ir}} - \frac{\sum_{i=1}^N Y_i (1-Z_i) w_{i0} S_{ir}}{\sum_{i=1}^N (1-Z_i) w_{i0} S_{ir}}. \tag{6}$$

In this paper, we focus on two weighting schemes, the inverse probability weighting:[21] $(w_{i1}, w_{i0}) = \left(\frac{1}{e(X_i, S_i)}, \frac{1}{1-e(X_i, S_i)}\right)$; and the overlap weighting:[17] $(w_{i1}, w_{i0}) = (1 - e(X_i, S_i), e(X_i, S_i))$. Both inverse probability weighting and overlap weighting estimators are unbiased for subgroup average treatment effect due to randomization.[14, 26]

To implement propensity score weighting method for subgroup analysis in RCTs, one could fit a propensity score model. If the model is specified by logistic regression,

$$\hat{e}(X_i, S_i) = logit^{-1}(\hat{\alpha}_0 + X_i^T \hat{\alpha}_X + S_i^T \hat{\alpha}_S + (X_i S_i)^T \hat{\alpha}_{XS}), \tag{7}$$

where $(\hat{\alpha}_0, \hat{\alpha}_X^T, \hat{\alpha}_S^T, \hat{\alpha}_{XS}^T)^T$ are the maximum likelihood estimators. When $S_i$ is a function of $X_i$, or is not included in the set of $X_i$, then $(X_i S_i)$ should include the full interactions between $S_i$ and all adjustment covariates $X_i$. When $S_i$ is a subset of $X_i$, then $(X_i S_i)$ should include the full interactions between $S_i$ and the rest of the covariates in the adjustment set. Yang et al.[29] showed that overlap weighting exactly balances the mean of the covariates with interaction terms in each subgroup as well as the overall sample,

$$\sum_{i=1}^{N} Z_i S_{ir} X_{ip} \hat{w}_{i1} - \sum_{i=1}^{N}(1 - Z_i) S_{ir} X_{ip} \hat{w}_{i0} = 0, \tag{8}$$

for each $r = 1, \ldots, R$, and $p = 1, \ldots, P$. This property has important practical implications in subgroup analyses. For any baseline covariate-subgroup interaction included in the propensity score model, the associated chance imbalance in the subgroup sample and in the overall sample of a specific randomized trial vanishes once the overlap weights are applied. Moreover, the exact balance property can translate into better efficiency in estimating subgroup average treatment effect with a finite sample. In contrast, this property does not generally hold for inverse probability weighting. Intuitively, the estimation error of an unadjusted estimator can be decomposed into the covariate imbalance term and the random error term. The estimation error of overlap weights is free of chance imbalance and only depends on the random error term, which equals to the estimation error when a true outcome model is fit (see Supplementary Material B for technical details about potential efficiency gain). This smallest estimation error for each treatment allocation translates into larger efficiency over repeated experiments. Extending the work of Zeng et al.[16] to subgroups, we can show that under some mild conditions, the class of propensity score weighting estimators is asymptotically equivalent to the "ANCOVA-S" estimator $\hat{\tau}_r^{ANCOVA}$ for estimating subgroup average treatment effect in RCTs.

In RCTs, the consistency of the $\hat{\tau}_r^h$ in (6) does not depend on the specification of the propensity score model. This is because any propensity score model leads to a consistent estimate of $e(X_i, S_i)$, which equals the randomization probability in a specific RCT. Shen et al.[15] and Williamson et al.[22] have provided analytical arguments to show the variance reduction properties of inverse probability weighting for estimating the average treatment effect in RCTs. Overlap weighting has similar results, but showed improved finite-sample performance over inverse probability weighting.[16] By similar arguments, we can show that the propensity score weighting estimator improves the precision of the unadjusted estimator, $\hat{\tau}_r^{UNADJ}$, if the propensity score model adjusts for important prognostic covariates. Variance and confidence intervals of the $\hat{\tau}_r^h$ can be obtained either via M-estimation-based sandwich estimator[18, 30-32] or bootstrap resampling.

**Simulation Study**

We consider a variety of simulation scenarios to compare the performance of the unadjusted estimator, the ANCOVA-S, the inverse probability weighting and overlap weighting estimators with different propensity score model specifications. Because all estimators above for subgroup average treatment effect are essentially unbiased, we evaluate these methods based on relative efficiency, coverage rate, type I error rate and power for the associated hypothesis testing. Consistent with the recent RCTs published in top medical journals, we generate $N \in \{250, 500, 750, 1000\}$ patients to represent small, moderate and large trials.[2] In each data set, subgroup indicators $\boldsymbol{S}_i = (S_{i1}, \ldots, S_{iR})^T$ are independently drawn from Bernoulli(0.25), resulting in $S_{ir} = 1$ being the smaller subgroup level. We generate $J$ independent adjustment variables $\boldsymbol{X}_i = (X_{i1}, \ldots, X_{iJ})^T$, half of which are continuous, and the other half are binary. The continuous covariates are independently drawn from the standard normal distribution, and the binary covariates are drawn from Bernoulli(0.3). The randomized treatment assignment $Z_i$ is generated from Bernoulli(0.5). Although under this setting, the true propensity score model is known, the outcome model is not known and may be complex. We simulate from two outcome models to represent different complexity levels. In outcome model 1, the continuous outcome $Y_i$ is generated from a normal distribution with mean $E(Y|Z, \boldsymbol{S}, \boldsymbol{X})$ specified below and standard deviation $\sigma_y = 1$, where fit the ANCOVA-S model would be appropriate:

$$E(Y|Z, \boldsymbol{S}, \boldsymbol{X}) = \beta_0 + \boldsymbol{\beta}_1^T \boldsymbol{X} + \boldsymbol{\beta}_2^T \boldsymbol{S} + \beta_3 Z + \boldsymbol{\beta}_4^T (\boldsymbol{XS}) + \boldsymbol{\beta}_5^T (Z\boldsymbol{S}) \\ + \boldsymbol{\beta}_6^T (\boldsymbol{X}Z) + \boldsymbol{\beta}_7^T (\boldsymbol{XS}Z). \tag{9}$$

In outcome model 2, to consider the possibility that the data generating model maybe more complex than fit by ANCOVA-S, we add some arbitrary interactions. Specifically, $Y_i$ additionally depend on the $(J-1)$ interactions between the pairs of covariates with consecutive indices, denoted by $\boldsymbol{X}_{int} = (X_1 X_2, \ldots, X_{J-1} X_J)^T$:

$$E(Y|Z, \boldsymbol{S}, \boldsymbol{X}) = \beta_0 + \boldsymbol{\beta}_1^T \boldsymbol{X} + \boldsymbol{\beta}_2^T \boldsymbol{S} + \beta_3 Z + \boldsymbol{\beta}_4^T (\boldsymbol{XS}) + \boldsymbol{\beta}_5^T (Z\boldsymbol{S}) \\ + \boldsymbol{\beta}_6^T (\boldsymbol{X}Z) + \boldsymbol{\beta}_7^T (\boldsymbol{XS}Z) + \boldsymbol{\beta}_8^T \boldsymbol{X}_{int}. \tag{10}$$

We refer to this scenario as "outcome model misspecification" henceforth. Details of parameter specifications in the two outcome models are given in Supplementary Material C.

For propensity score weighting estimators, the propensity score model specification is key. Because including subgroup-covariate interactions in the propensity score model leads to improved balance of covariates within subgroups, and is necessary to achieve exact subgroup covariate balance via overlap weighting,[29] we consider two types of logistic propensity score models. The first is the main-effects logistic model, which includes the subgroup indicator and the covariate main effects; the corresponding design matrix is $(X, S)$. This model represents the standard practice in the literature. The second is a full-interaction model, which, besides the subgroup variables and the main effects, also includes their 2-way interactions; the corresponding design matrix is $(X, S, XS)$. Given the estimated propensity score, we use the Hájek-type estimator (6) for both inverse probability weighting and overlap weighting. For the ANCOVA-S estimator, we fit the ANCOVA model in (9) (outcome model 1) and estimate subgroup average treatment effect using $\hat{\tau}_r^{ANCOVA}$ in (5). In total four simulation scenarios are summarized in Table 1. Additional details of each scenario, together with variance estimation and evaluation metrics are provided in the Supplementary Material C.

**Scenario 1**

In Scenario 1, a single baseline covariate is used to create subgroups (e.g. a high-risk co-morbidity; $R = 1$),[1, 6] and the treatment effect is additive and homogeneous within subgroups. We set $J = 8$ to represent a common number of covariates adjusted in the RCTs,[1] and vary $(\beta_3, \beta_5) \in \{(0,0), (-1, 0.5)\}$, representing the treatment effects are null or subgroup-treatment interaction is half as the main treatment effect. Next, we set $\boldsymbol{\beta_6}, \boldsymbol{\beta_7}$ to be vector of zeros, representing homogeneous treatment effect within subgroups (i.e. $\tau_{\{S=0\}} = \beta_3, \tau_{\{S=1\}} = \beta_3 + \beta_5$).

When testing the null hypothesis of no treatment effect or no heterogeneous treatment effect across subgroup levels, the rejection rates under the null hypothesis and coverage rates under the alternative are close to the nominal 5% and 95% levels (Figure 1(a) and Supplementary Table 2). In Figure 2(a), the efficiency of all adjusted estimators is higher than the unadjusted estimator. Specifically, the efficiency of propensity score weighting with full interaction model has the same efficiency as ANCOVA-S. When subgroup sample size is relatively small (e.g. <125), the overlap weighting estimator with the propensity scores estimated from the full-interaction model (OW-Full) leads to higher efficiency compared to ANCOVA-S and the inverse probability weighting estimator with a full-interaction propensity model (IPW-Full). In contrast, when the propensity scores are estimated from the main-effect model, both inverse probability weighting (IPW-Main) and overlap weighting (OW-Main) perform poorly in estimating subgroup average treatment effect. Figure 3(a) displays the results on power analysis. For testing subgroup average treatment effect in the smaller subgroup ($S = 1$), all methods are comparable, but ANCOVA-S and IPW-Full have smaller power when sample size is less than 100. For testing the heterogeneous treatment effect, OW-Full leads to higher power when subgroup sample size is less than 125.

**Scenario 2**

Scenario 2 differs from scenario 1 by allowing for heterogeneous treatment effect within subgroups. To do this, we fix $(\beta_3, \beta_5) = (-1, 0.5)$, $\boldsymbol{\beta_6} = 0.5 \times \mathbf{1_8}$, and $\boldsymbol{\beta_7} = 0.25 \times \mathbf{1_8}$, where $\mathbf{1_k}$ is a $k$-vector of ones. The choice of $\boldsymbol{\beta_7}$ represents the interaction effect size is half as the subgroup-treatment effect $\beta_5$. By numerical integration over the true model, $\tau_{\{S=0\}} = -0.4, \tau_{\{S=1\}} = 0.4$. The coverage rates of all estimators are close to the nominal level

(Supplementary Table 3). Similar to Scenario 1, when subgroup sample size is less than 125, the efficiency of OW-Full is comparable to the ANCOVA-S, followed by the IPW-Full. The IPW-Main and OW-Main perform equally poor (Supplementary Figure 1(a)). The result of power confirms this trend. OW-Full and ANCOVA-S estimator have consistently higher power compared to other estimators, followed by the IPW-Full and main effect estimators. The unadjusted estimator has the lowest power throughout (Supplementary Figure 1(b)).

**Scenario 3**

Under the similar framework to Scenario 1, we explore the performance of various estimators under outcome model misspecification. The outcomes are generated from model (10) with $\boldsymbol{\beta_8} = \sqrt{\sigma_y/(J-1)}$ (1,1,1,1,1,1,1), representing effects of the covariate interactions. Other regression coefficients remain the same as in Scenario 1. Again, the rejection rates under the null hypothesis and coverage rates under the alternative are close to the nominal levels (Figure 1(b) and Supplementary Table 4). Figure 2(b) demonstrates that when subgroup sample size is less than 125, OW-Full and ANCOVA-S estimator are comparable, both outperforming IPW-Full estimator, and the main effect estimators. OW-Full becomes slightly more efficient than ANCOVA-S when the outcome model is misspecified. Figure 3(b) shows when testing subgroup average treatment effect in the smaller subgroup ($S = 1$), OW-Full estimator has the highest power. ANCOVA-S and IPW-Full have even lower power than unadjusted and the main effect estimators when subgroup sample size is less than 100, but outperform them when subgroup sample size is larger than 125. For testing the heterogeneous treatment effect, OW-Full estimator leads to consistently higher power than all other estimators.

**Scenario 4**

Scenario 4 extends Scenario 1 to study multiple pre-specified subgroups of interest ($R = 6$; 12 subgroups), corresponding to the number of subgroups being analyzed in our real data example. We vary the number of subgroup variables with non-zero coefficients $M \in \{1, 4\}$ (i.e. whether the elements of $\boldsymbol{\beta_2}, \boldsymbol{\beta_4}, \boldsymbol{\beta_5}$ are non-zeros, $\boldsymbol{\tau}_{\{S=0\}} = \beta_3 \times \mathbf{1}_R, \boldsymbol{\tau}_{\{S=1\}} = \beta_3 + \boldsymbol{\beta_5}$). We compare a joint approach which includes all pairwise interactions of $(\boldsymbol{X}, \boldsymbol{S}, \boldsymbol{XS})$ in Equations (7) and (4) with a one-at-a-time approach (i.e. iterative design matrix $(\boldsymbol{X}, \boldsymbol{S_r}, \boldsymbol{XS_r})$ for $r = 1, ..., R$). The one-at-a-time approach is motivated by the fact that the joint approach may be infeasible when the number of subgroups and covariate combinations is large relative to the sample size. Supplementary Figure 2 shows the one-at-a-time approach performs similarly with the joint approach, and slightly better when the subgroups are not prognostic of the outcome. As with all multiple testing problems, the family-wise error rate is inflated for all methods. Full description and results of Scenario 4 are provided in the Supplementary Material C and Supplementary Figure 2-4, 5(a).

**Summary of Simulation Findings**

In summary, across all the scenarios, efficiency of the adjusted estimators is higher than that of the unadjusted estimator. For the weighting estimators, the full-interaction propensity model consistently outperforms the main-effect propensity model. For large subgroups, overlap weighting and inverse probability weighting perform similarly, but overlap weighting is more efficient than inverse probability weighting in small samples (e.g. <125). When the ANCOVA model is correctly specified, weighting methods with full-interaction is as efficient as

ANCOVA-S, and is even more efficient when paired with overlap weighting under small subgroup sample size (e.g. <125). When the ANCOVA model is misspecified, OW-Full outperforms the ANCOVA-S estimator. Lastly, in the presence of multiple subgroups, one-at-a-time analysis performs similarly to the joint approach, and the family-wise error rate is inflated for all methods.

We also investigate scenarios with multiple subgroups that have heterogeneous treatment effect within subgroups, misspecified outcome model, and both heterogeneous treatment effect and misspecified outcome model (Supplementary Table 1). The results on relative efficiency and power are similar to those of the single subgroup scenarios, and the family-wise error rate of ANCOVA-S from the misspecified model may be larger than that of the propensity score weighting estimators (Supplementary Figure 5b, 6-8).

**Application to the HF-ACTION Trial**

The Heart Failure: A Controlled Trial Investigating Outcomes of Exercise Training (HF-ACTION) is a patient-level randomized trial that evaluated the efficacy and safety of an exercise training program in chronic heart failure and reduced ejection fraction patients (N=2331).[23] We evaluated a cohort of 1724 patients with the outcome of 6 minute walk distance (6MWD) measured at 3 months. The investigators pre-specified subgroups (prior to databaseline lock) based on 10 baseline covariates. Figure 4 shows a Connect-S plot[29] of representative subgroups that have imbalanced covariates between treatment arms. Each row represents subgroups of interest, and each column represents adjustment variables. The color of the dots represents severity of covariate imbalanced, measured by the absolute standardized mean difference. While the traditional IPW-Main results in severe imbalance (absolute standardized mean difference greater than 0.2) in subgroups defined by angina classification, ventricular conduction, angiotensin-converting enzyme (ACE) inhibitors and $\beta$-blocker usage at baseline, the OW-Full achieves exact balance in all subgroups. For illustration, we estimate the subgroup average treatment effect in these imbalanced subgroups adjusting four baseline covariates: baseline 6-minute walk distance, age, race, and duration of the cardiopulmonary exercise test. These adjustment covariates were selected for having a strong relationship to outcome based on clinical knowledge and prior publications. Although the data were complete for most candidate covariates, <1% patients are missing baseline 6-minute walk distance or cardiopulmonary exercise test. Missing data was handled with multiple imputation, with 10 imputed data sets derived using R package *mice*.

The subgroups are evaluated one-at-a-time, in the sense that $R = 1$ in the preceding equations, and the methods are applied separately for each subgroup of interest. Within each imputed data set, the ANCOVA-S model includes a treatment indicator, a single subgroup variable, adjustment variables, and all 2-way interactions ($XZ, XS, ZS$). To estimate the propensity scores, we fit the full-interaction logistic model, which includes the single subgroup variable, adjustment variables, and their interactions ($X, S, XS$). Within-imputation variances are estimated via 1000 bootstrap samples. The final point and variance estimates are obtained by the Rubin's rule.[33]

Figure 5 presents the subgroup average treatment effect estimates, standard errors (SEs), and 95% confidence intervals (CIs) of unadjusted, IPW-Full, OW-Full, and ANCOVA-S. In all subgroups, SE from the unadjusted estimator is larger than the adjusted estimators, and three adjusted estimators have comparable SEs. The point estimate from the unadjusted estimator is slightly larger in most subgroups. It is worth mentioning that for testing subgroup average treatment effect in four subgroups (delayed intraventricular conduction, angina class I, angina class II-IV and no $\beta$-blocker usage), the difference between methods may alter the conclusion about subgroup treatment effect because the unadjusted estimate includes the null, while the adjusted methods do not. Specifically, results from unadjusted estimate suggest that there is no statistically significant effect of exercise training on 6-minute walk distance at 3 months in these subgroups. In contrast, results from the adjusted methods suggest that treated patients on average walk 25, 23, 31, 22 meters longer in 6-minute walk distance than control patients in these four subgroups, separately. This discrepancy highlights the potentially significant efficiency gain from covariate adjustment in subgroup analysis.

**Discussion**

In this paper, we extend ANCOVA and introduce two propensity score weighting methods (inverse probability weighting and overlap weighting) for covariate adjustment in subgroup analyses of RCTs. Through extensive simulations, we found that for propensity score weighting it is crucial to include covariates predictive of the outcome, the subgroup variable of interest and its interactions with covariates. When the subgroup sample size is large, overlap weighting and inverse probability weighting perform similarly, but overlap weighting consistently outperforms inverse probability weighting in terms of power with small subgroups (e.g. <125). Finally, OW-Full outperforms the ANCOVA-S estimator under small subgroup sample size (e.g. < 125) and/or when the ANCOVA model is misspecified. This is because the exact balance property of overlap weighting guarantees all covariates are balanced within small subgroups, which translates into improved power in estimating the subgroup average treatment effect.

In practice, these methods are designed for the setting where important covariates and subgroup variables are pre-specified based on clinical knowledge. Alternative approaches to search for heterogeneous treatment effect across covariates using machine learning models have been proposed elsewhere,[35-38] but their utility in RCT remains to be shown. In addition, one should weigh the benefit of uncovering a finer resolution of treatment effect heterogeneity with the potential risk of over-fitting with a finite sample size. As the number of subgroup variables and covariates increases, the number of their pairwise interactions would grow exponentially, leading to non-convergence in model fitting and unstable propensity score estimates. Variable selection in propensity score model has been discussed in the context of subgroup analyses in observational study.[29] Here, where the true propensity score is known to be null, we found the one-at-a-time approach performs similarly to the joint approach. However, the joint approach is more efficient when a subgroup is highly prognostic of outcome, but not already included as a covariate, because it incorporates all of the subgroups at once. This illustrates the general principle that ANCOVA will be more efficient if it includes factors that are predictive of the outcome. In practice, it is prudent to pre-choose adjustment variables that are the most prognostic of the outcome. In the HF-ACTION example, the number of pre-specified subgroups is large, so we adopt the one-at-a-time approach, which could balance covariates within subgroup as well as avoiding over-complicated models.

Last, although we focused on a continuous outcome, the propensity score weighting estimators are directly applicable to RCTs with binary and categorical outcomes. In this case, the propensity score weighting methods preserve the definition of subgroup estimands averaged over pre-treatment covariates within the subgroup and may be preferred, for the same reason that weighting is preferred in estimating the overall ATE in RCTs.[15, 16, 22]

We provide publicly available R and SAS functions for implementing methods used in this paper at: https://github.com/siyunyang/SGA_RCT.

(a)

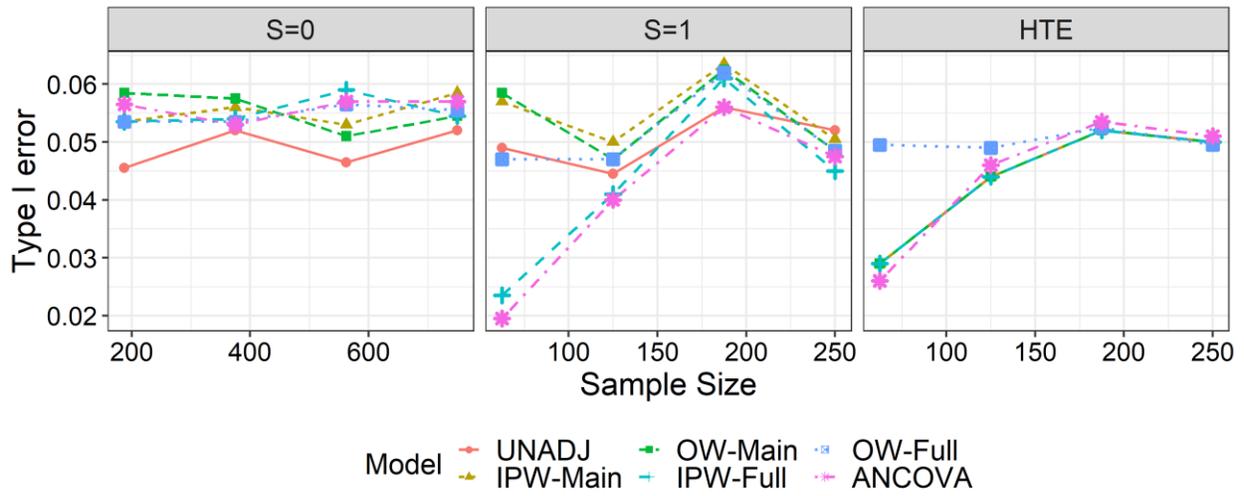

(b)

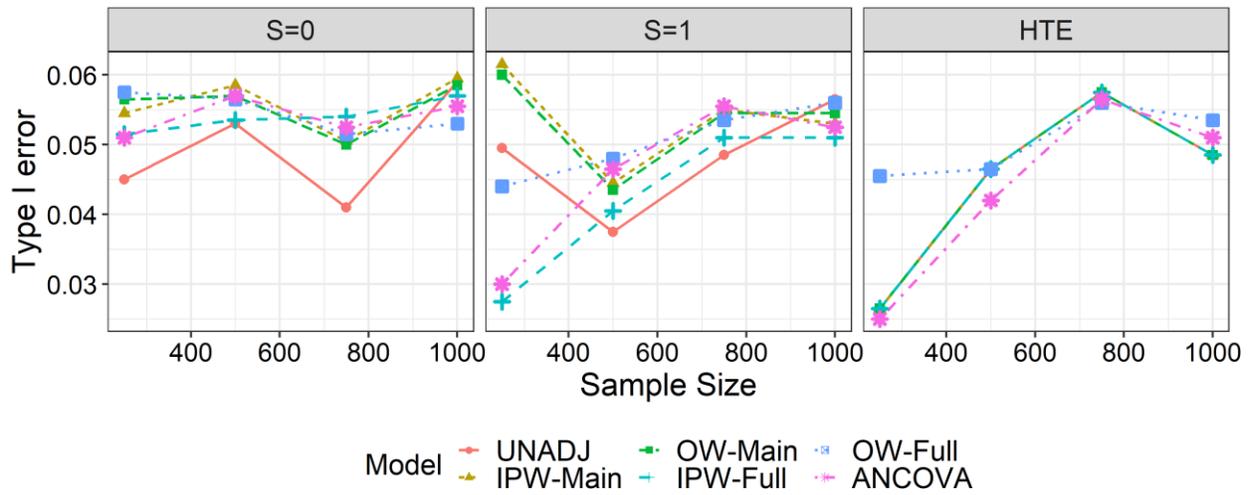

**Figure 1.** The Type I error rate of different estimators for testing no treatment effect of subgroup average treatment effects and heterogeneous treatment effect (HTE) across subgroups under (a) Scenario 1; (b) Scenario 3.

(a)

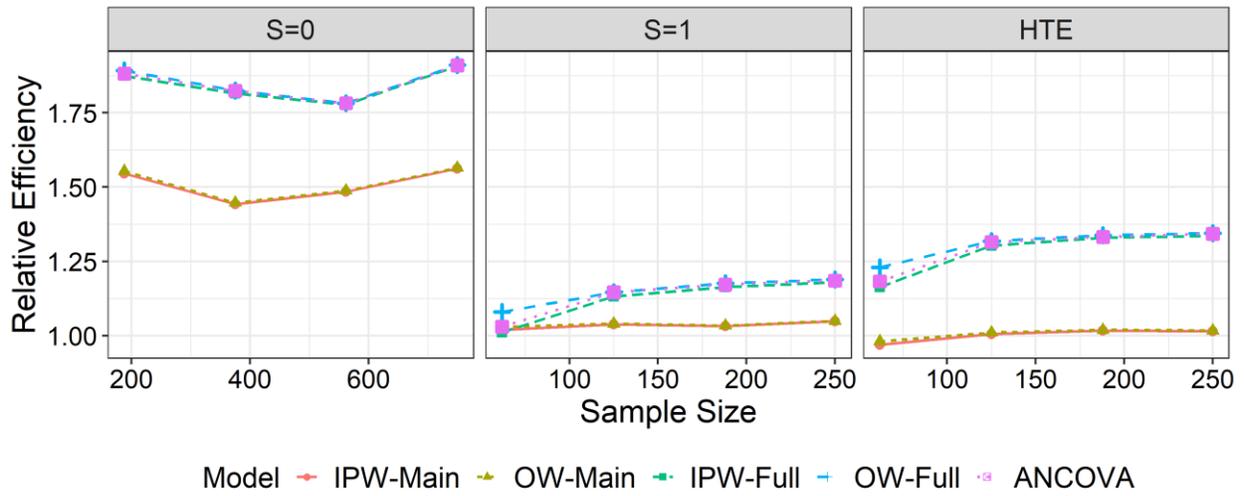

(b)

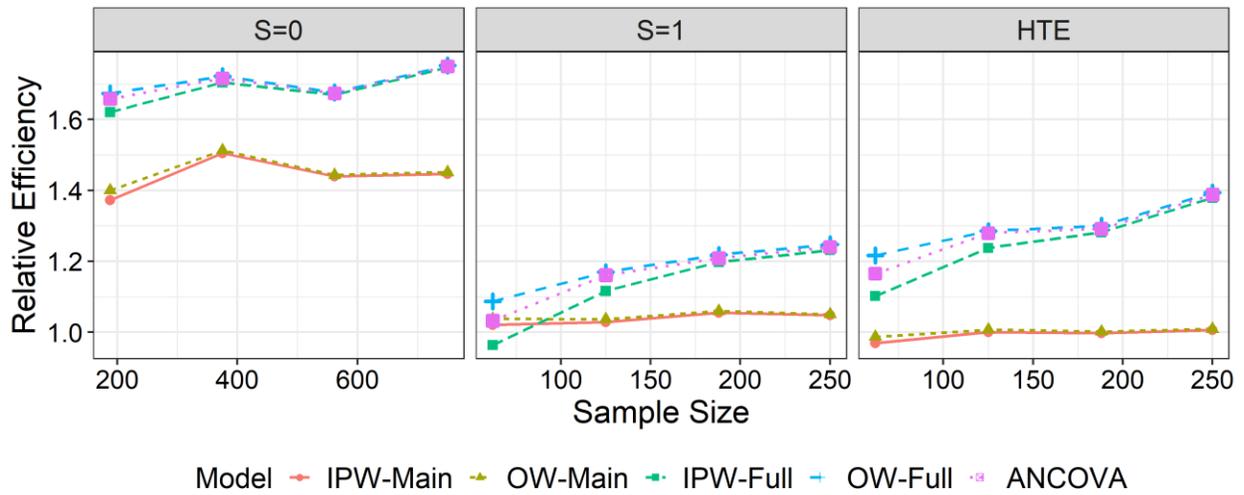

**Figure 2.** The relative efficiency of adjusted estimators relative to unadjusted estimator for estimating subgroup average treatment effects and heterogeneous treatment effect (HTE) across subgroups when $\beta_3 = -1, \beta_5 = 0.5$ under (a) Scenario 1, $\beta_6 = \beta_7 = \vec{0}$ when ANCOVA model is correctly specified; (b) Scenario 3, $\beta_6 = \beta_7 = \mathbf{0}$ when ANCOVA model is misspecified.

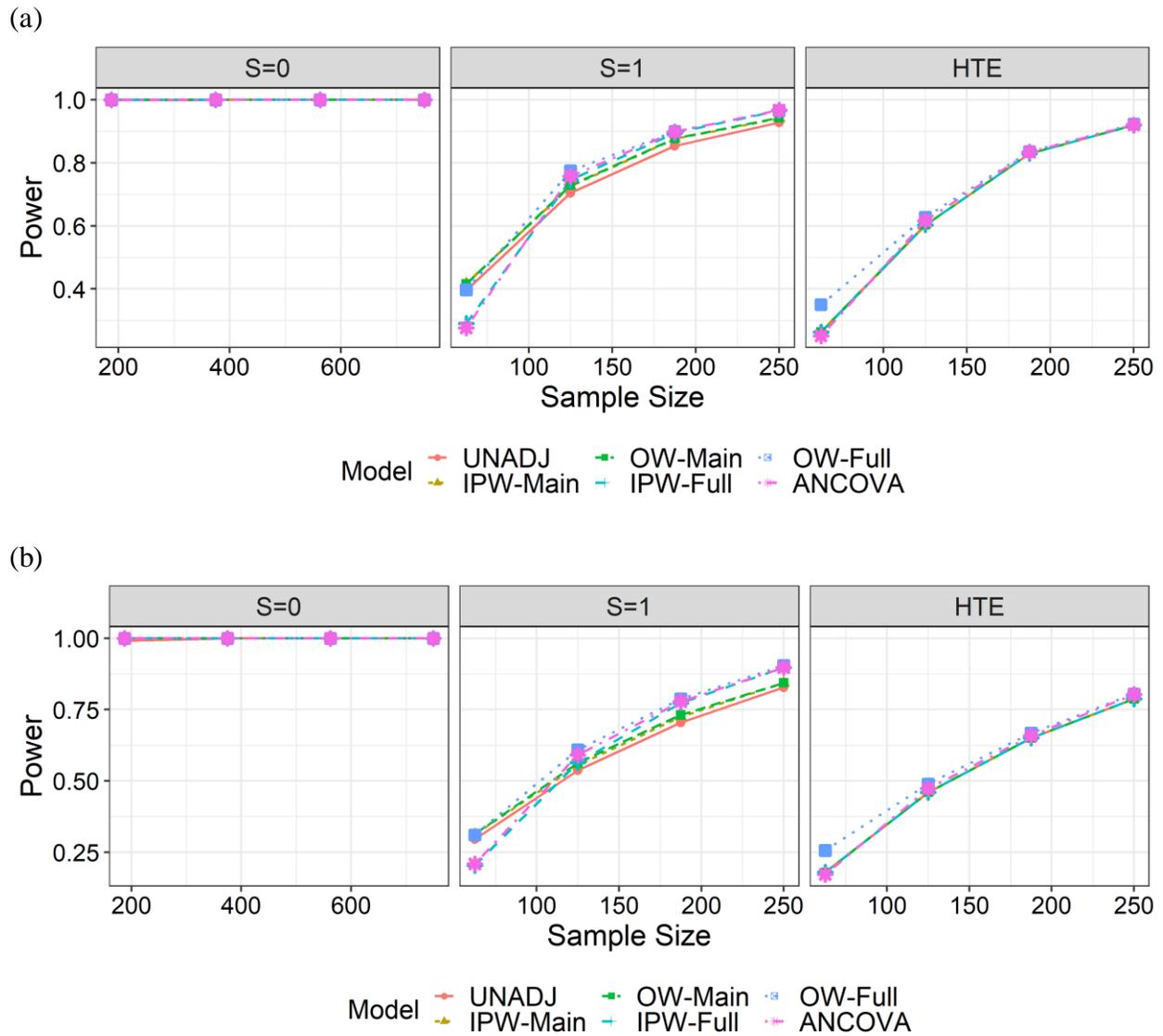

**Figure 3.** The power of different estimators for estimating subgroup average treatment effects and heterogeneous treatment effect (HTE) across subgroups when $\beta_3 = -1, \beta_5 = 0.5$ under (a) Scenario 1, $\beta_6 = \beta_7 = 0$ when ANCOVA model is correctly specified; (b) Scenario 3, $\beta_6 = \beta_7 = 0$ when ANCOVA model is misspecified.

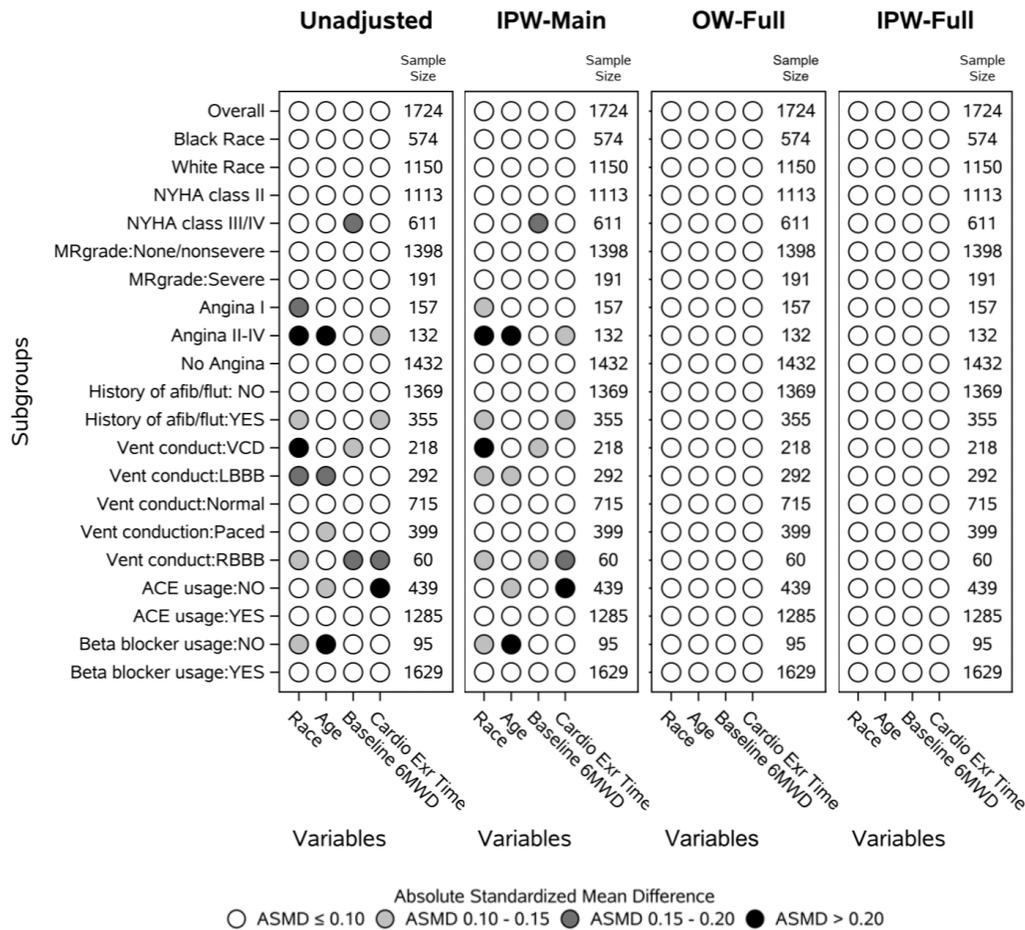

**Figure 4.** The Connect-S plot of subgroup absolute standardized mean difference in HF-ACTION trial. Each row represents subgroups of interest, and each column represents adjustment variables. The color of the dots represents severity of covariate imbalanced, measured by the absolute standardized mean difference. NYHA indicates New York Heart Association; MRgrade, mitral regurgitation grade; Angina, Canadian Cardiovascular Society Angina class; afib/flut: atrial fibrillation or atrial flutter; Vent Conduction, ventricular conduction; VCD, intraventricular conduction delay; LBBB, left bundle branch block; RBBB, right bundle branch block; ACE: angiotensin-converting enzyme; 6MWD: 6-minute walk distance.

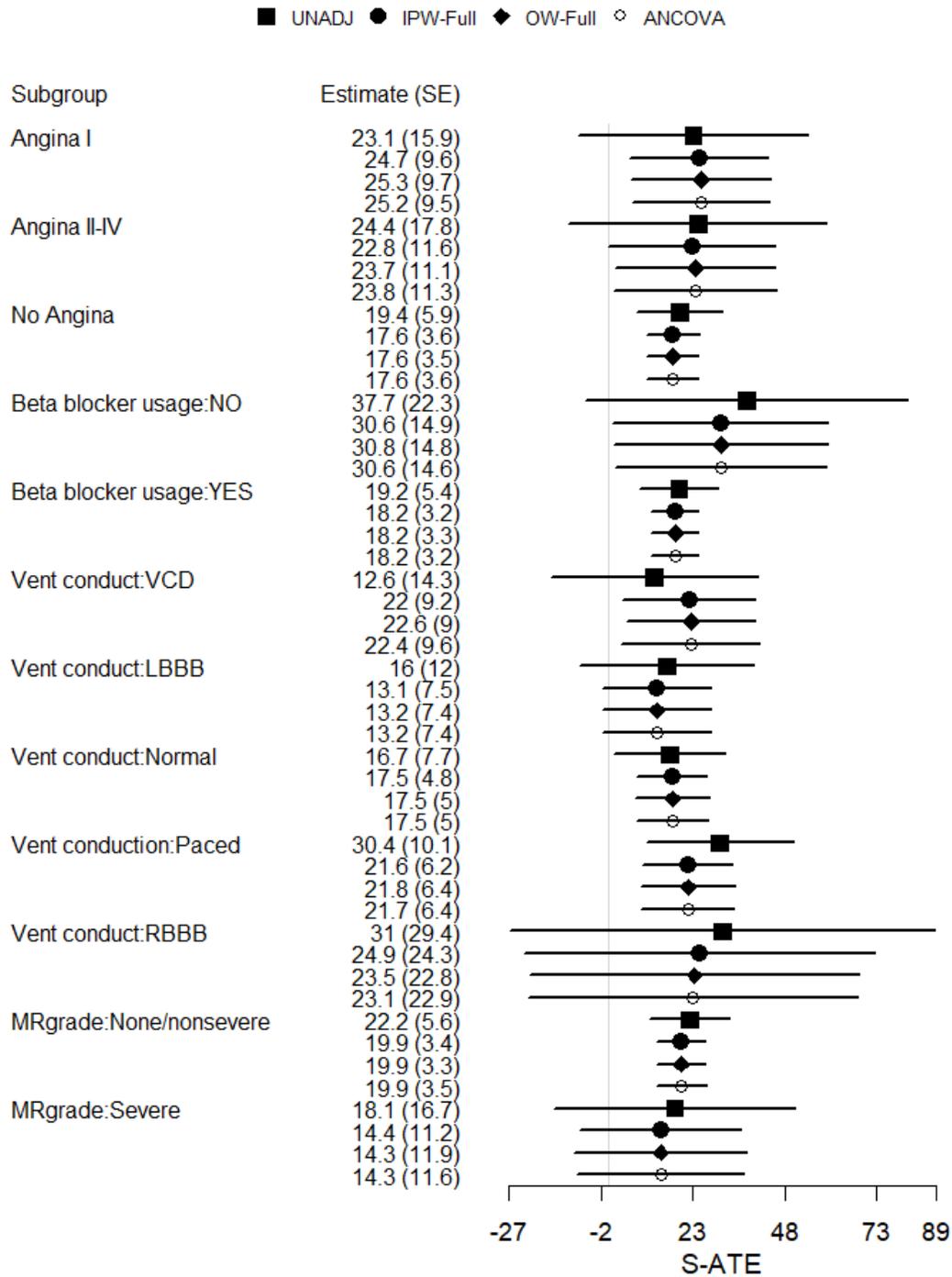

**Figure 5.** Subgroup average treatment effect (S-ATE) estimates on 6-minute walk distance at 3 months from the HF-ACTION study.

**Table 1.** Summary of simulation scenarios.

| | Heterogeneous treatment effect within subgroups | | Outcome model misspecification | | Subgroup variables | |
|---|---|---|---|---|---|---|
| | Homogeneous | Heterogeneous | Yes | No | Single | Multiple |
| Scenario 1 | x | | | x | x | |
| Scenario 2 | | x | | x | x | |
| Scenario 3 | x | | x | | x | |
| Scenario 4 | x | | | x | | x |